\documentclass[10pt,fleqn]{article}
\usepackage{wrapfig}
\usepackage{amsbsy,epsf,amssymb}
\usepackage{multicol}
\newcommand{\bk}{\boldsymbol {k}}
\newcommand{\bq}{\boldsymbol {q}}

\begin{document}
%
%
\begin{center}
{\bf\Large  Self-consistent approach for thermodynamics of a
            simplified pseudospin-electron model}
\end{center}
%
%
\centerline{ I.V.~Stasyuk, A.M.~Shvaika, and K.V.~Tabunshchyk}
%
%
\begin{center}
 Institute for Condensed Matter Physics Nat. Acad. Sci. Ukr.\\
            1 Svientsitskii Str., UA--79011 Lviv, Ukraine
\end{center}

\begin{multicols}{2}
{\small
 We present the method of the self-consistent calculation of
thermodynamical and correlation functions. This approach is based
on the GRPA (generalized random phase approximation) scheme with
the inclusion of the mean field corrections.
 Investigation of a pseudospin-electron model (PEM) within the framework
of the presented method shows that interaction between the
electron and pseudospin subsystems leads to the possibility of
either first or second order phase transitions between different
uniform phases (bistability) as well as between the uniform and
the chess-board ones.
 In the regime $n={\rm const}$, an instability with respect to phase
separation in the electron and pseudospin subsystems can take
place.\\
 $$\vspace{-.7cm}$$
}

\section{Introduction}

 The crystals with the high-temperature superconductivity (HTSC)
investigated intensively during the last ten years possess a wide
spectrum of the interesting physical properties.
 The variety of effects which are realized only separately in the
case of other types of crystals is characteristic of these
systems.
 The strong electron correlation, related to the interaction of the
Hubbard type in the conducting bands formed mainly by the
superconducting Cu$_{2}$-O$_{2}$ plains, can be pointed out as one
of the reasons of this unique situation.

 Another significant feature connected with the dynamics of HTSC crystals is the
presence of strongly anharmonic elements of the structure.
 As it is known, it can be a source of instabilities of various type.

 Among the most frequently investigated HTSC crystals are
the crystals of the YBa$_{2}$Cu$_{3}$O$_{7-\delta}$ group.
 The unit cell of these compounds contains, besides two superconducting
planes, the chain (at the $\delta\ll 1$ composition) elements
Cu$_{1}$-O$_{1}$, connected by Cu$_{1}$-O$_{4}$-Cu$_{2}$ bridges
with conducting plains through the apical oxygen ions O$_{4}$.
 The vibrations of these ions along the $c$-axis (perpendicularly to the plains)
exhibit a strong anharmonicity.
 Much evidence exists in support of this concept.
 One can mention the results of experimental investigations
(EXAFS data [1,2], Raman scattering \mbox{[3-6]}, dielectric
measurements [7,8]) where the two equilibrium positions of O$_{4}$
ion (two different values of the $R_{\rm O_{4}-Cu_{2}}$ distance)
have been observed that can point out to the presence of the local
double-minimum potential well.
 Similar conclusions were made  based on  consideration  the local polaron
phenomena [9,10], electron transfer processes through O$_{4}$ ions
[11,12] or bistabilities in the normal phase temperature region
[13].

 Besides, a connection between positions of O$_{4}$ ions and
the energy of electron states in Cu$_{2}$-O$_{2}$ plains plays an
important role in YBa$_{2}$Cu$_{3}$O$_{7-\delta}$ crystals.
 The data given in [14] point to the existence of a
significant correlation between the occupancy of electron states
of the Cu$_{2}$ ion and the R$_{\rm O_{4}-Cu_{2}}$ distance as
well as to the decrease of this distance at the transition from
the metallic orthorombic phase to the semiconducting phase (that
takes place at $\delta >\delta^{*}=0.55$).
 These and other similar facts suggest the presence of a large
electron-vibrational coupling.

 By now, the description of the locally anharmonic subsystem in
HTSC crystals develops with the use of two different approaches.
 The first one, which is chronologically older, is based on the
phonon anharmonic model $\varphi^{4}$ [15], or, more recently, on
the model $\varphi^{3}+\varphi^{4}$ [13].
 This approach was used while consideration the polaron effect [16]
and also while investigating the effect of anharmonicity  on the
superconducting transition temperature and on the possibility of
the modulated (of the CDW type) phase creation [17].

 In the second approach, which is more appropriate at the strong
el\-ec\-tron-vib\-ra\-ti\-on\-al coupling and in the cases when
two equilibrium positions of anharmonic ions really exist, the
pseudospin formalism is used; the pseudospin variable
$S_{i}^{z}=\pm 1/2$ defines these two positions.
 This scheme being applied to the YBa$_{2}$Cu$_{3}$O$_{7-\delta}$
type systems started from works [18] and [19].

 On the basis of the second model which was called as
pseudospin-electron model (PEM) a possible connection between the
superconductivity and lattice instability of the ferroelectric
type in HTSC has been discussed [19,20].
 The description of the electron spectrum and the electron statistics of the PEM
was given in [21] in the framework of the temperature two-time
Green function method in the Hubbard-I approximation.

 A series of works has been carried out in which the pseudospin
$\left\langle S^zS^z\right\rangle$, mixed $\left\langle
S^zn\right\rangle$ and charge  $\left\langle nn\right\rangle $
correlation functions were calculated.
 It has been shown with the use of the generalized random phase
approximation (GRPA) [22] in the limit of infinite single-site
electron correlations ($U\to\infty$) [23,24], that there exists a
possibility of divergences of these functions at some values of
temperature.
 This effect was interpreted as a manifestation of dielectric instability or
ferroelectric type anomaly.
 The tendency to the spatially modulated charge and pseudospin ordering
at the certain model parameter values was found out.

 On the other hand, the case of absence of the term describing
electron transfer in Cu$_{2}$-O$_{2}$ plains ($t_{ij}=0$) with the
inclusion of the direct interaction between pseudospins was
considered within the mean field approximation [25,26] (see also
the short review in [27]).
 The first or second order phase transitions
with the jumps of $\langle S^z\rangle$ and electron concentration
$n$ values in the $\mu={\rm const}$ regime were obtained. An
instability with respect to phase separation in the electron and
pseudospin subsystems can take place in regime $n={\rm const}$.

 The analysis of ferroelectric type instabilities in the two-sublattice
model of the apex oxygen subsystem in high temperature
superconducting systems has been made [28].
 The influence of oxygen nonstoichiometry on localization of apex
oxygen in YBa$_2$Cu$_3$O$_{7-\delta}$ type crystals was studied in
the work [29].

 In the present work we propose the self-consistent scheme for calculation
of mean values of pseudospin and electron number operators, grand
canonical potential as well as correlation functions for the case
of the $U=0$ limit (the simplified PEM).
 The approach is based on the GRPA with the inclusion of the mean
field type contributions coming from effective pseudospin
interactions via conducting electrons [30].
 The main attention is paid to the thermodynamics of phase transitions.
 The possibilities of phase separation and chess-board phase
appearance are investigated.

\section{Pseudospin-electron model}

Hamiltonian of the pseudospin-electron model has the following
form:
\begin{eqnarray}
\hspace{-1.3cm}
 &&H{=}H_{0}+\sum_{ij\sigma} t_{ij}b_{i\sigma}^{+} b_{j\sigma},\nonumber\\
 \hspace{-1.3cm}
 &&H_{0}{=}\sum_{i}
   \left\{Un_{i\uparrow}n_{i\downarrow} {-} \mu\sum_{\sigma}n_{i\sigma}
   {+} g\sum_{\sigma} n_{i\sigma} S_{i}^{z} {-} hS_{i}^{z} \right\},\nonumber
 \label{Hamiltonian1}
\end{eqnarray}
where the strong single-site electron correlation $U$, interaction
with the anharmonic mode ($g$-term) and the energy of the
anharmonic potential asymmetry ($h$-term) are included in the
single-site part; $\mu$ is the chemical potential.
 The second term in the Hamiltonian describes the electron hopping from site to
site (the electron transfer parameter $t_{ij}$).

 The formalism of electron annihilation (creation) operators
$
 a_{i\sigma}=b_{i\sigma}P^+_i,
$
$
 \tilde{a}_{i\sigma}=b_{i\sigma}P^-_i
$
($P^{\pm}_i=\frac 12\pm S^z_i$) acting at a site with the certain
pseudospin orientation is introduced.
 The calculation is performed in the strong coupling case ($g\gg t$) using
of single-site states as the basic one.
\begin{eqnarray}
\label{Hamiltonian2}
 \hspace{-1.3cm}
 &&H_0{=}\sum\limits_i\{\varepsilon
 (n_{i\uparrow}{+}n_{i\downarrow})+
 \tilde{\varepsilon}(\tilde{n}_{i\uparrow}{+}\tilde{n}_{i\downarrow})-
 hS^z_i\},\\
 \hspace{-1.3cm}
 &&H_{\rm int}{=}\sum\limits_{ij\sigma}t_{ij}
 (a^+_{i\sigma}a_{j\sigma}{+}
  a^+_{i\sigma}\tilde{a}_{j\sigma}{+}
 \tilde{a}^+_{i\sigma}a_{j\sigma}{+}
 \tilde{a}^+_{i\sigma}
 \tilde{a}_{j\sigma}).\nonumber
\end{eqnarray}

Here $\varepsilon =-\mu+g/2 ,\quad \tilde{\varepsilon}
=-\mu-g/2\quad$ are energies of the single--site states.
 We consider here the simplified PEM (the case $U=0$).

 Expansion of the calculated quantities in terms of electron transfer
leads to the infinite series of terms containing the averages of
the $T$-products of the $a_{i\sigma}$, $\tilde{a}_{i\sigma}$
operators.
 The evaluation of such averages is made using the corresponding
Wick's theorem.
 The results are expressed in terms of the products
of nonperturbed Green functions and averages of a certain number
of the projection operators $P^{\pm}_i$ which are calculated by means
of the semi-invariant expansion [30].

 Single-electron Green function (calculated in Hub\-bard-I type approximation)
may be written as the following chain diagram:
\begin{equation}
\hspace{-.8 cm}
 \label{Pgreen} \raisebox{-.7cm}{\epsfysize 1.1cm\epsfbox{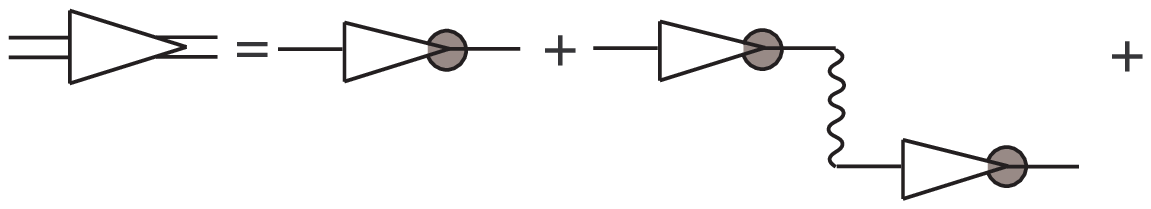}}
 \quad \dots\quad .
\end{equation}

 In the adopted approximation the diagrammatic series for the pseudospin mean
value can be presented in the form
\begin{equation}
 \label{Sz}
 \hspace{-.94cm}
 \langle S^z_i\rangle{=} \raisebox{-.23cm}{\epsfysize 1.4cm\epsfbox{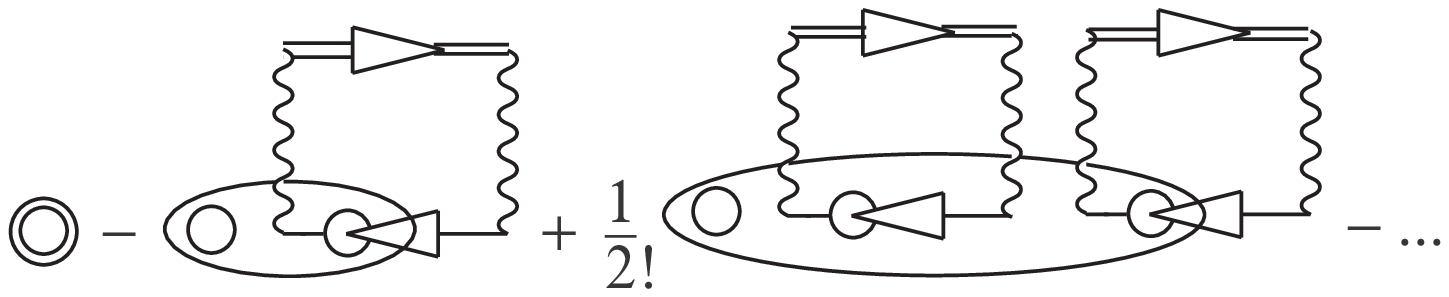}}\,.
\end{equation}
 Here we use the following diagrammatic notations:
$
 \raisebox{-.13cm}{\epsfysize .4cm\epsfbox{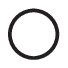}}
 {=}S^z_i
$,
$
 \raisebox{-.13cm}{\epsfysize .4cm\epsfbox{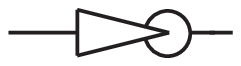}}
$
$
\displaystyle
 {=} g_i(\omega_n)
 {=}\frac{P^+_i}{{\rm i}\omega_n-\varepsilon}+
 \frac{P^-_i}{{\rm i}\omega_n-\tilde{\varepsilon}},
$
 nonperturbated electron Green function
$
  \raisebox{-0.13cm}{\epsfysize .4cm\epsfbox{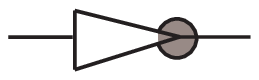}}
 {=} \langle g_i(\omega_n)\rangle,
$
wavy line is the intersite hopping $t_{ij}$.
 Semi-invariants are represented by ovals and contain the
$\delta$-symbols on site indexes.
 In the spirit of the traditional mean field approach [30] the
renormalization of the basic semi-invariant by the insertion of
independent loop fragments is taken into account in (\ref{Sz}).

 The  diagrammatic series for the electron concentration mean
value is the following:
\begin{equation}
\label{n}
 \hspace{-1cm}
 \langle n_i\rangle {=}
\raisebox{-1.61cm}{\epsfysize 2.8cm\epsfbox{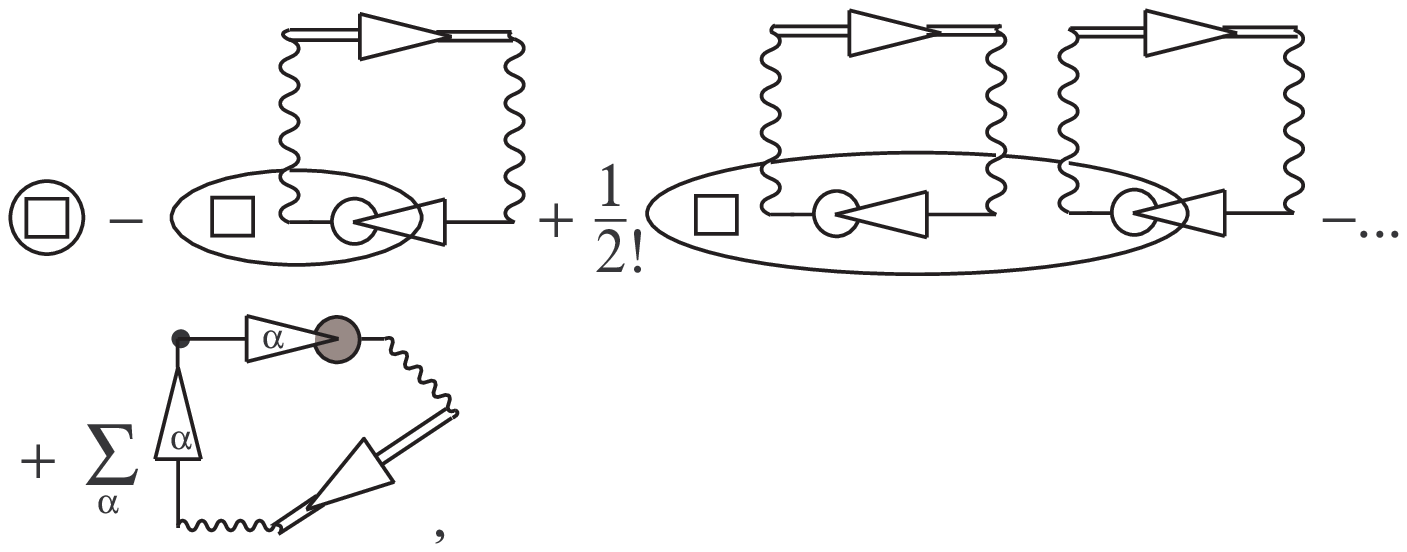}}
\end{equation}
where
$
 \raisebox{-0.1cm}{\epsfysize .4cm\epsfbox{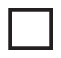}}{=}n_i,
$
$
\displaystyle
 \raisebox{-0.1cm}{\epsfysize .4cm\epsfbox{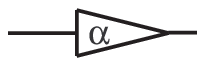}}
 {=} \frac1{{\rm i}\omega_n{-}\varepsilon^{\alpha}},
$
$
\displaystyle
 \raisebox{-0.1cm}{\epsfysize .4cm\epsfbox{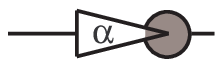}}
 {=}\frac{\langle P^{\alpha}_i\rangle}{{\rm i}
  \omega_n-\varepsilon^{\alpha}},
$
$
 \varepsilon^{\alpha}{=}(\varepsilon,\tilde{\varepsilon}),
$
$
 P^{\alpha}_i{=}(P^+_i,P^-_i).
$

 In the same approximation the grand canonical potential
and pseudospin correlation function are, respectively:
\begin{equation}
  \label{Gcp}
  \hspace{-.7cm}
  \Delta\Omega=\hspace*{-1.3cm}
  \raisebox{-2.3cm}{\epsfysize 3.5cm\epsfbox{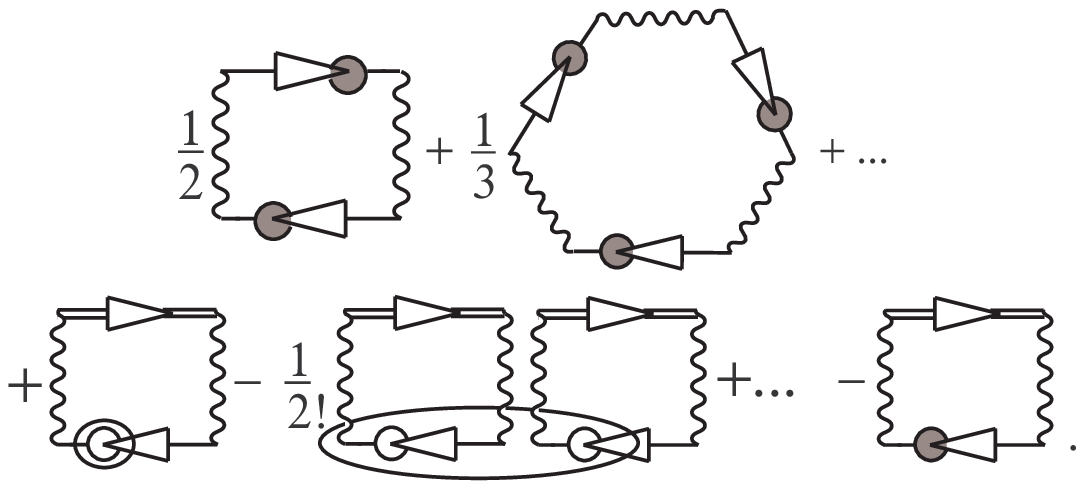}}
  \hspace{-2cm}
\end{equation}
\begin{equation}
  \label{SzSz}
  \hspace{-1cm}
  \langle S^z_iS^z_j\rangle=
  \raisebox{-2.15cm}{\epsfysize 2.6cm\epsfbox{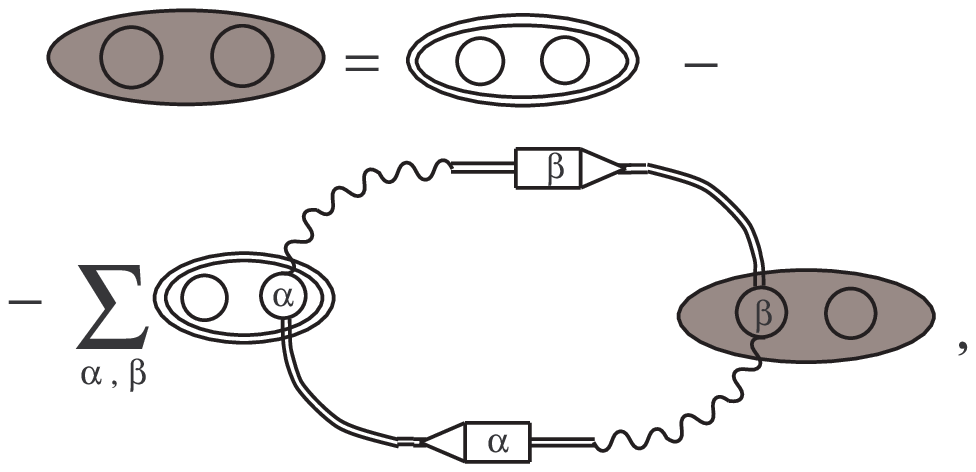}}
\end{equation}
$$
 \hspace{-.3cm}
 \raisebox{-.12cm}{\epsfysize .4cm\epsfbox{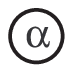}}
 =P^{\alpha}_i,\quad
 \raisebox{-1.cm}[.3cm][.9cm]{\epsfysize1.3cm\epsfbox{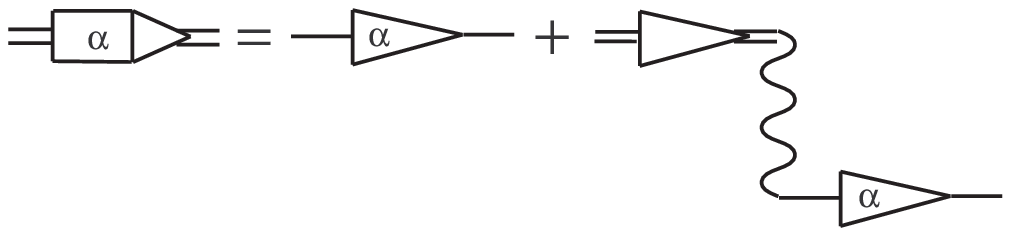}}.
$$
   First term in equation $(\ref{SzSz})$ takes into
account a direct influence of the internal effective
self-consistent field on pseudospins:
\begin{equation}
  \label{15}
   \hspace{-1.cm}
    \raisebox{-1.4cm}[1.2cm][1.2cm]{\epsfysize 2.8cm\epsfbox{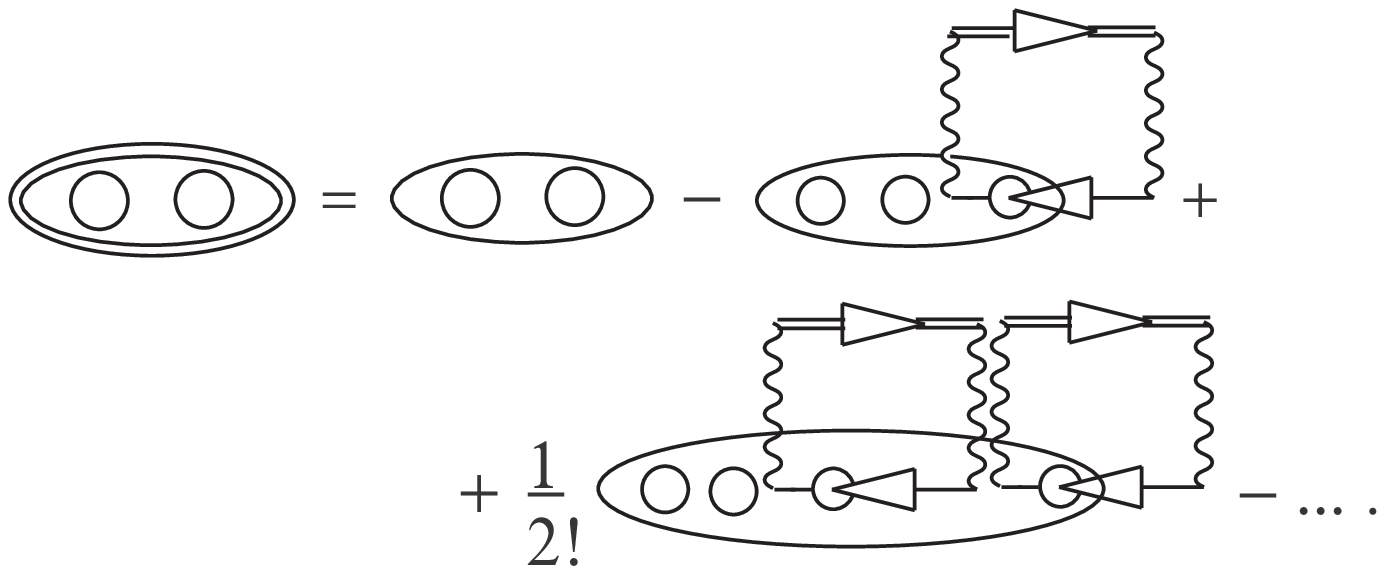}}
\end{equation}
leading to the renormalization  of the second-order semi-invariant
due to the inclusion of ``single-tail'' parts.
 Second term in equation $(\ref{SzSz})$ describes an interaction
between pseudospins which is mediated by electron hopping.

 Hence, with respect to the GRPA scheme an influence of the
internal effective self-consistent field on pseudospins is taken
into account by means of the inclusion of the mean field type
contributions into the expressions for all thermodynamic
quantities.
 In the approach presented the correlation functions are
calculated consistently with thermodynamical functions.

 The consistency of the expressions (\ref{Sz})--(\ref{SzSz}) can be checked
explicitly [30] using the thermodynamical relations:
$$
 \frac{{\rm d}\Omega}{{\rm d}(-\mu)}
 =\langle n\rangle,\quad
 \frac{{\rm d}\Omega}{{\rm d}(-h)}
 =\langle S^z\rangle,\quad
 \frac{{\rm d}\langle S^z\rangle}{{\rm d}(\beta h)}
 =\langle S^zS^z\rangle_{{\bq}=0}\;.
$$

 At high temperatures eq. (\ref{Sz}) possesses only the uniform
solution $\langle S^z_i\rangle =\langle S^z\rangle$.
 The phase transitions between uniform phases with different pseudospin
mean values $\langle S^z\rangle$ were analyzed in [30].
 For the first time the possibility of the dielectric
instabilities has been done for the PEM in the limit of the strong
electron correlation ($U\rightarrow\infty$) in [23,24].
 Results of ref. [30], where the opposite case of $U\to 0$ was analyzed,
 are in good agreement with the exact ones for the $U\to 0$ PEM
in the limit of infinite spatial dimension [31].
 At the same time, a complete description of such transitions was obtained in
[25,26] for the PEM with direct interaction between pseudospins
($t_{ij}=0$ limit).

 On the other hand, the solution of eq. (\ref{SzSz}) for pseudospin
correlator has the form

\begin{equation}
 \label{2.4}
 \hspace{-.7cm}
 \langle S^z S^z\rangle_{\bq}
   =\frac{1/4-\langle S^z\rangle^2}
   {1+\mbox{\fbox{$\Pi$}}_{\bq}(\frac{1}{4}-\langle S^z\rangle^2)},
\end{equation}
where \fbox{$\Pi$}$_{\bq}$ characterize an interaction between
pseudospins via electron subsystem:
\begin{equation}
 \label{petlia}
 \mbox{\fbox{$\Pi$}}_{\bq}=\sum\limits_{\alpha,\beta}(-1)^{\alpha+\beta}
 \raisebox{-.75cm}[.5cm][.5cm]{\epsfysize 1.5cm\epsfbox{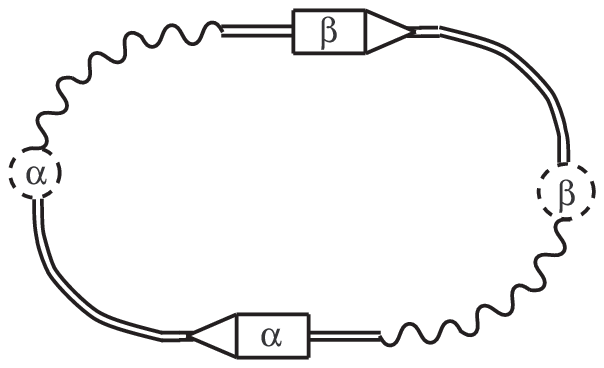}}\;\;,
\end{equation}
and its singularities
$$
 1+\mbox{\fbox{$\Pi$}}_{\bq}(\frac{1}{4}-\langle
 S^z\rangle^2)=0
$$
 gives the points of the instability of uniform
phase with respect to fluctuations with wave vector ${\bq}$.
 The typical dependence of the temperature of instability on the
fluctuation wave vector ${\bq}$ for square lattice is shown in
figure~1 and one can see that for some model parameter values
uniform phase become unstable with respect to fluctuations with
${\bq}=(\pi,\pi)$ (chess-board phase).
\begin{center}
 \epsfxsize 0.4\textwidth\epsfbox{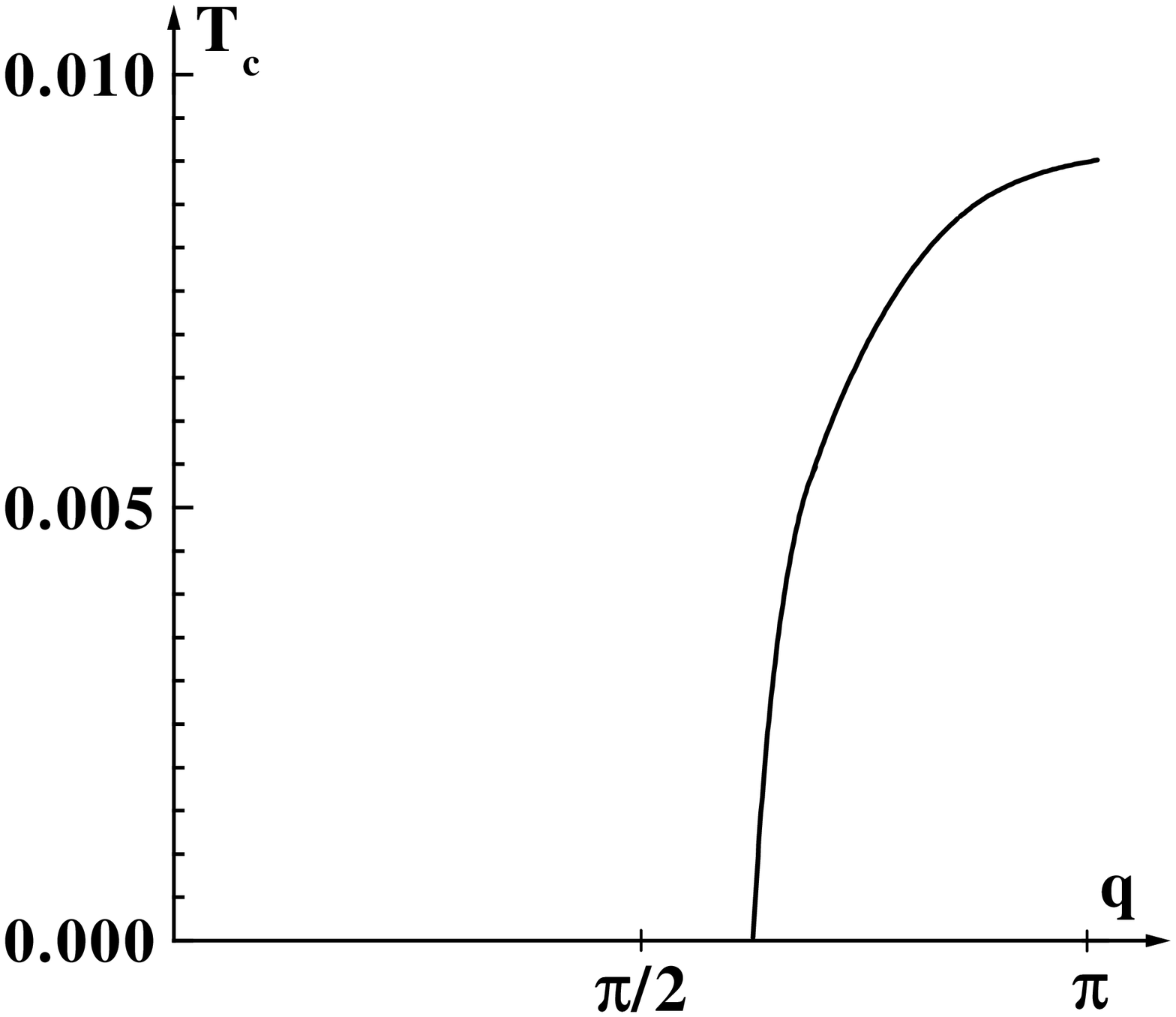}
\end{center}
 {\small Figure~1. Temperature of the uniform phase instability versus the
fluctuation wave vector ${\bq}=(q,q)$ for square lattice
 ($h=0.272$, $\mu=-0.36$, $W=0.2$, $g=1$, the lattice
spacing is seted to 1); $W$ is the halfwidth of the initial
electron band.}
 $$\vspace{-.4cm}$$
 So, below we shall take into account the possibility of the chess-board
phase appearance in our analysis.

\section{The chess-board phase}

 To take into account the modulation of the pseudospin and electron
distribution we introduce two kinds of sites: the $\langle
S^z_1\rangle$ corresponds to the one sublattice, and the $\langle
S^z_2\rangle$ to the other one.
 The nearest--neighbour hopping exists only between the sublattices.

 Single--electron Green functions (\ref{Pgreen}) in this case are
equal:
\begin{eqnarray}
\label{G1G2}
 G_{1\bk}(\omega_n)&=&\frac{g_1(\omega_n)}
 {1-t^2_{\bk}g_1(\omega_n)g_2(\omega_n)},\quad \\
 G_{2\bk}(\omega_n)&=&\frac{g_2(\omega_n)}
 {1-t^2_{\bk}g_1(\omega_n)g_2(\omega_n)},\nonumber
\end{eqnarray}
where $g_1(\omega_n)$ and $g_2(\omega_n)$ are nonperturbated Green
functions for the sublattices 1 and 2 respectively.
 The poles of functions $G_{\bk}(\omega_n)$ determine the
single--electron  spectrum. The equation for the spectrum has the
form:
\begin{eqnarray}
\label{eq2}
 &&\hspace*{-1cm}x^4-(g^2/2+t_{\bk }^2)x^2- gt_{\bk }^2(\langle
 S^z_1\rangle+\langle S^z_2\rangle)x+\\
 &&\hspace*{-.5cm}+g^4/16-g^2t_{\bk }^2\langle
 S^z_1\rangle\langle S^z_2\rangle=0,\nonumber
\end{eqnarray}
where $x={\rm i}\omega_n+\mu$.

 The roots
$
 \varepsilon_1(t_{\bk })\geqslant
 \varepsilon_2(t_{\bk })\geqslant
 \varepsilon_3(t_{\bk })\geqslant
 \varepsilon_4(t_{\bk })
$
of the equation (\ref{eq2}) form four subbands.
 The widths of subbands depend on the mean values of pseudospins.

 The branches $\varepsilon_1(t_{\bk })$, $\varepsilon_2(t_{\bk })$ on the one
side and $\varepsilon_3(t_{\bk })$, $\varepsilon_4(t_{\bk })$ on
the other one coincide at $t_{\bk }=0$
 ($\varepsilon_{1,2}(t_{\bk }=0)=g/2$,
 $\varepsilon_{3,4}(t_{\bk}=0)=-g/2$)
and form two pairs of bands which are always separated by gap.

 The equations for pseudospin mean values (\ref{Sz}) can be
written now in the form:
\begin{equation}
 \label{Szeq2}
  \hspace{-.8cm}
 \langle S^z_l\rangle=
 \frac{1}{2}\tanh\left\{\frac{\beta}{2}(h{+}\alpha^l_2{-}\alpha^l_1)+
 \ln{\frac{1{+}{\rm e}^{-\beta\varepsilon}}
 {1{+}{\rm e}^{-\beta\tilde{\varepsilon}}}} \right\},
\end{equation}
 $l=1,2;\quad$
where expressions for the effective self-con\-sis\-tent fields are
\begin{equation}
  \hspace{-.7cm}
  \alpha^l_2{-}\alpha^l_1{=}
 \frac 2N\sum_{\bk}t^2_{\bk}(\varepsilon-\tilde{\varepsilon})
 \sum^4_{i=1}
  A^l_in[\varepsilon_i(t_{\bk}){-}\mu],
\end{equation}
$$
 \hspace{-.2cm}
 A^l_i{=}\frac{\varepsilon_i(t_{\bk})+g\langle S^z_{l'}\rangle}
{(\varepsilon_i(t_{\bk}){-}\varepsilon_j(t_{\bk}))
(\varepsilon_i(t_{\bk}){-}\varepsilon_p(t_{\bk}))
(\varepsilon_i(t_{\bk}){-}\varepsilon_m(t_{\bk}))},
$$
 $i\not=j,p,m,\quad l\not=l'.$\\
 Expression for electron number mean value follows from (\ref{n}):
\begin{eqnarray}
 \label{neq}
 \hspace*{-1.5cm}
 \lefteqn{\langle n_1{+}n_2\rangle{=}\frac{2}{N} \sum_{\bk}
 \sum^4_{i=1}n[\varepsilon_i(t_{\bk}){-}\mu]-}
 \\
 \nonumber
&& -2\Big[
  \big(\langle P_1^+\rangle{+}\langle P_2^+\rangle\big) n(\tilde{\varepsilon})
 +\big(\langle P_1^-\rangle{+}\langle P_2^-\rangle\big)
 n(\varepsilon)\Big],
\end{eqnarray}
and grand canonical potential (\ref{Gcp}) can be written in the
analytical form, also:
\begin{eqnarray}
 \label{Gcpeq}
 &&\hspace*{-1.5cm}
 \Delta\Omega =-\frac{2}{N\beta}\sum_{\bk}
 \ln\frac{
\prod\limits^4_{i=1}
\cosh\big[\frac{\beta}{2}(\varepsilon_i(t_{\bk}){-}\mu)\big]}
 {(\cosh\frac{\beta}{2}\varepsilon)^2
 (\cosh\frac{\beta}{2}\tilde{\varepsilon})^2}+\nonumber\\
 &&\hspace*{-1.5cm}
 +\sum_{l=1,2}\langle S_l^z\rangle(\alpha^l_2-\alpha^l_1)+\\
 &&\hspace*{-1.5cm}{+}\sum_{l=1,2}\left[-\frac{1}{\beta}\ln\cosh
  \left\{\frac{\beta}{2}(h{+}\alpha^l_2{-}\alpha^l_1)
   +\ln\frac{1{+}{\rm e}^{-\beta\varepsilon}}
           {1{+}{\rm e}^{-\beta\tilde{\varepsilon}}}
  \right\}\right.{+}\nonumber\\
  &&\hspace*{-1.5cm}\left.{+}\frac{1}{\beta}\ln \cosh
 \left\{ \frac{\beta}{2}h
  +\ln\frac{1{+}{\rm e}^{-\beta\varepsilon}}
           {1{+}{\rm e}^{-\beta\tilde{\varepsilon}}}
 \right\}\right]. \nonumber
\end{eqnarray}

\section{Numerical results}

 In the investigation of equilibrium conditions we shall separate
two different regimes in which the system can exist:

 a) the $\mu={\rm const}$ regime; it is supposed that the electron
states of other structure elements, which are not included
explicitly into the PEM, play a role of a thermostat, that ensures
a constant value of the chemical potential $\mu$ (despite the
possible changes of temperature, field $h$ and other
characteristics of the model).
 In this case the minimum of the grand canonical
potential $\Omega$ is a condition of thermodynamical equilibrium;

 b) the regime $n={\rm const}$; this situation is more customary at
the consideration of electron systems and it means that the
chemical potential is now the function of $T$, $h$ etc. and
depends on the electron concentration.
 The minimum of the free
energy $F=\Omega + \mu N$ is the equilibrium condition in this
case.

\subsection{$\mu={\rm const}$ regime}

 For this regime the equilibrium is defined by the minimum condition of $\Omega$
(\ref{Gcpeq}) that form the equations for pseudospin mean values
(\ref{Szeq2}) and expression for electron concentration
(\ref{neq}).
 The calculated field dependences of $\langle S^z_1 - S^z_2\rangle$
(order parameter for chess-board phase) and grand canonical
potential (which are determined by the solutions of the eqs.
(\ref{Szeq2})) are presented in figure~2 for $g\gg W$ and low
temperature.
\begin{center}
\raisebox{-2cm}[2cm][2cm]{\epsfysize 3.85cm\epsfbox{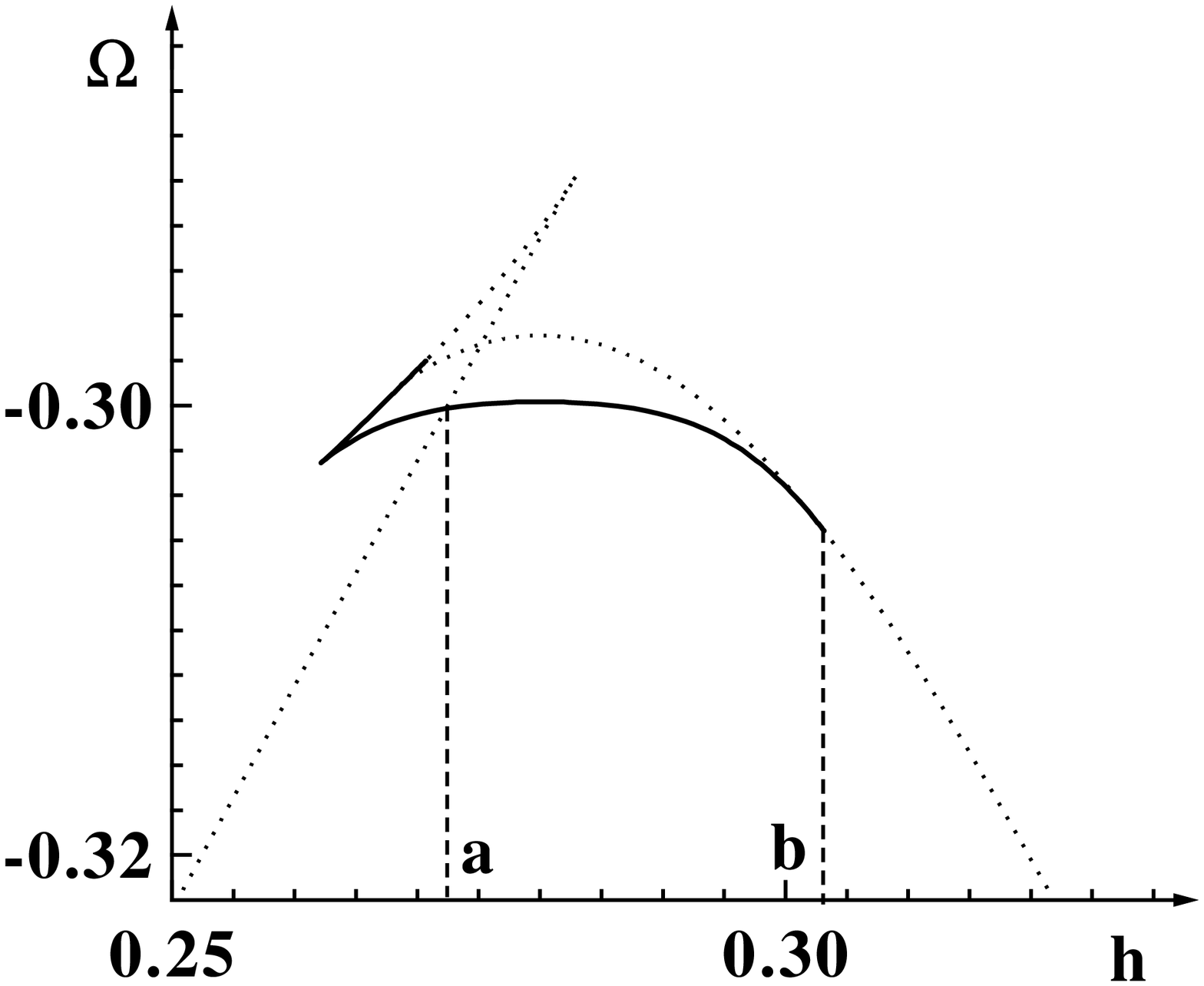}$\!$
                          \epsfysize 3.85cm\epsfbox{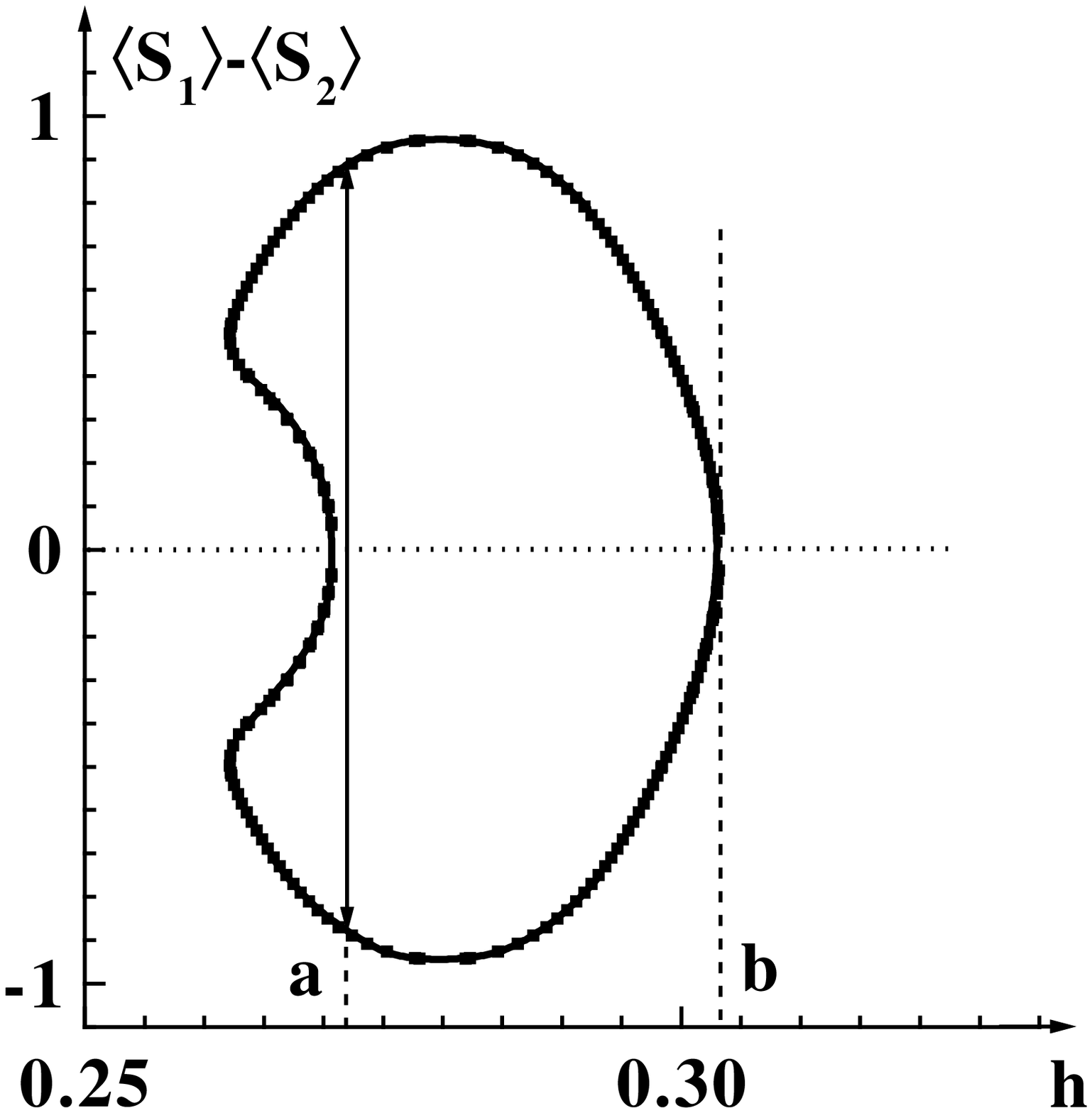}}
\end{center}
 {\small  Figure~2. Field dependence of the grand canonical
potential and order parameter ($T=0.005$, $\mu =-0.36$, $W=0.2$,
$g=1$).
 Dotted and solid lines correspond to the uniform and chess-board
phases, respectively.}
 $$\vspace{-.8cm}$$

 Comparison of the grand canonical potential $\Omega$
values for uniform and chess-board phases leads to the conclusion
that the modulated phase is thermodynamically stable at
intermediate values of $h$ parameter in the region between points
{\bf a} and {\bf b}.
 These points correspond to the first and second
order phase transition, respectively.
 The rapid jump--like change of order parameters is
accompanied by the rapid changes of the subbands widths and, as a
result, electron concentration.

 The resulting phase diagram $\mu-h$ at the low temperature is shown in figure~3.
\begin{center}
 \epsfxsize 0.4\textwidth\epsfbox{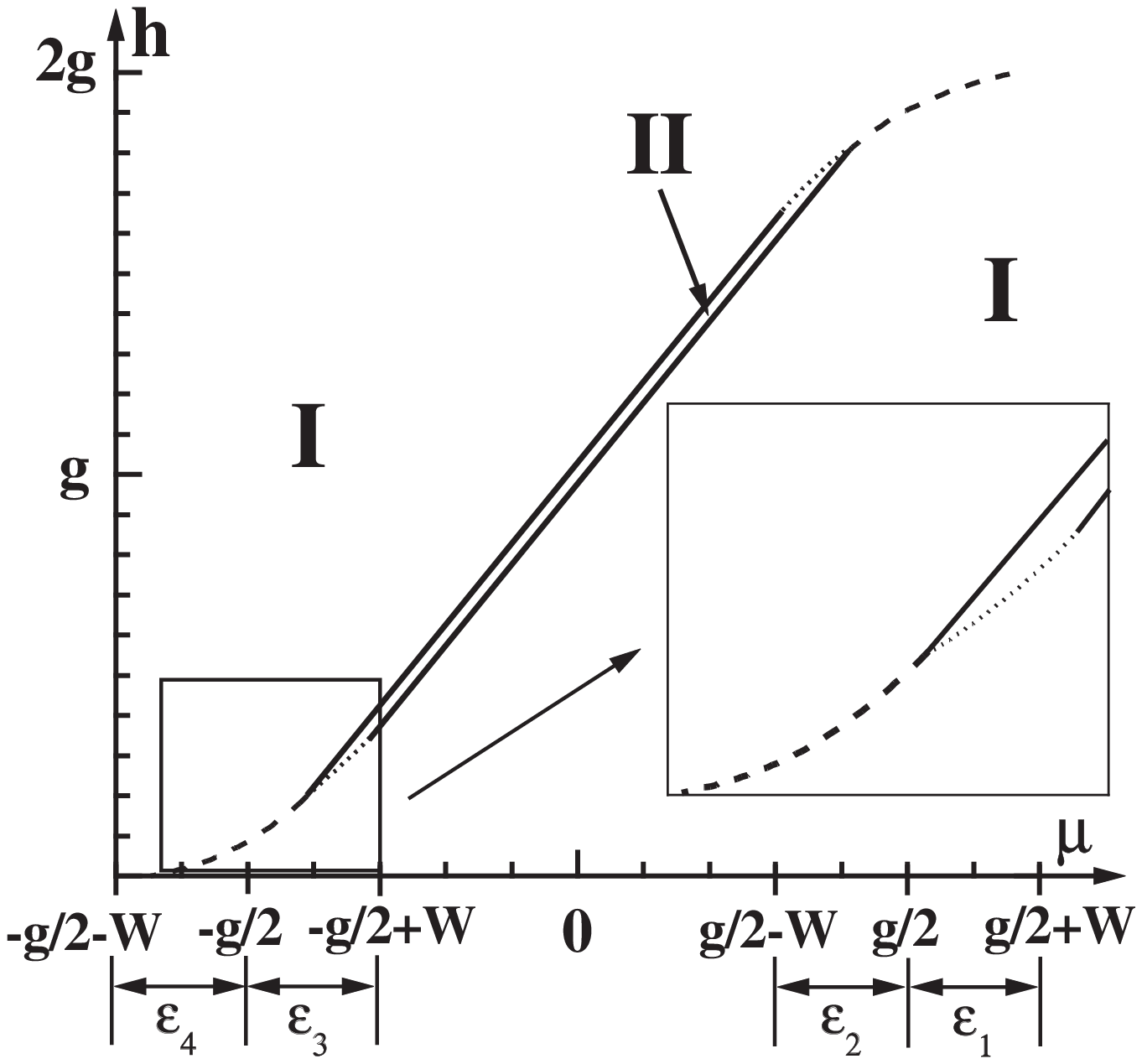}
\end{center}
 {\small  Figure~3. $\mu-h$ phase diagram
($T=0.005$, $W=0.2$, $g=1$). I -- uniform phase, II - chess-board
phase.
 Dashed lines -- first order phase transitions between the
uniform phases with different pseudospin mean values.
 Dotted lines -- first order phase transitions between the
uniform and chess-board phases.
 Solid lines -- second order phase transitions.}
$$\vspace{-.8cm}$$

 The transitions between uniform phases with different pseudospin
mean values as well as corresponding electron concentrations
(bistability), which is of the first order, takes place when the
chemical potential $\mu$ is placed within the
$
 \varepsilon_1
$,
$
 \varepsilon_2
$
and partially
$
 \varepsilon_3
$,
$
 \varepsilon_4
$
bands.
 The transitions between the uniform and modulated phases are of
the first or second order and can be realized in the case when
$\mu$ is placed in
$
 \varepsilon_2
$ and
$
 \varepsilon_3
$
bands or between them.
 The chess-board phase exists as intermediate
one between the uniform phases with different $\langle S^z\rangle$
and $n$ values.

 The $T-h$ phase diagram is shown in figure~4.
\begin{center}
 \epsfxsize 0.4\textwidth\epsfbox{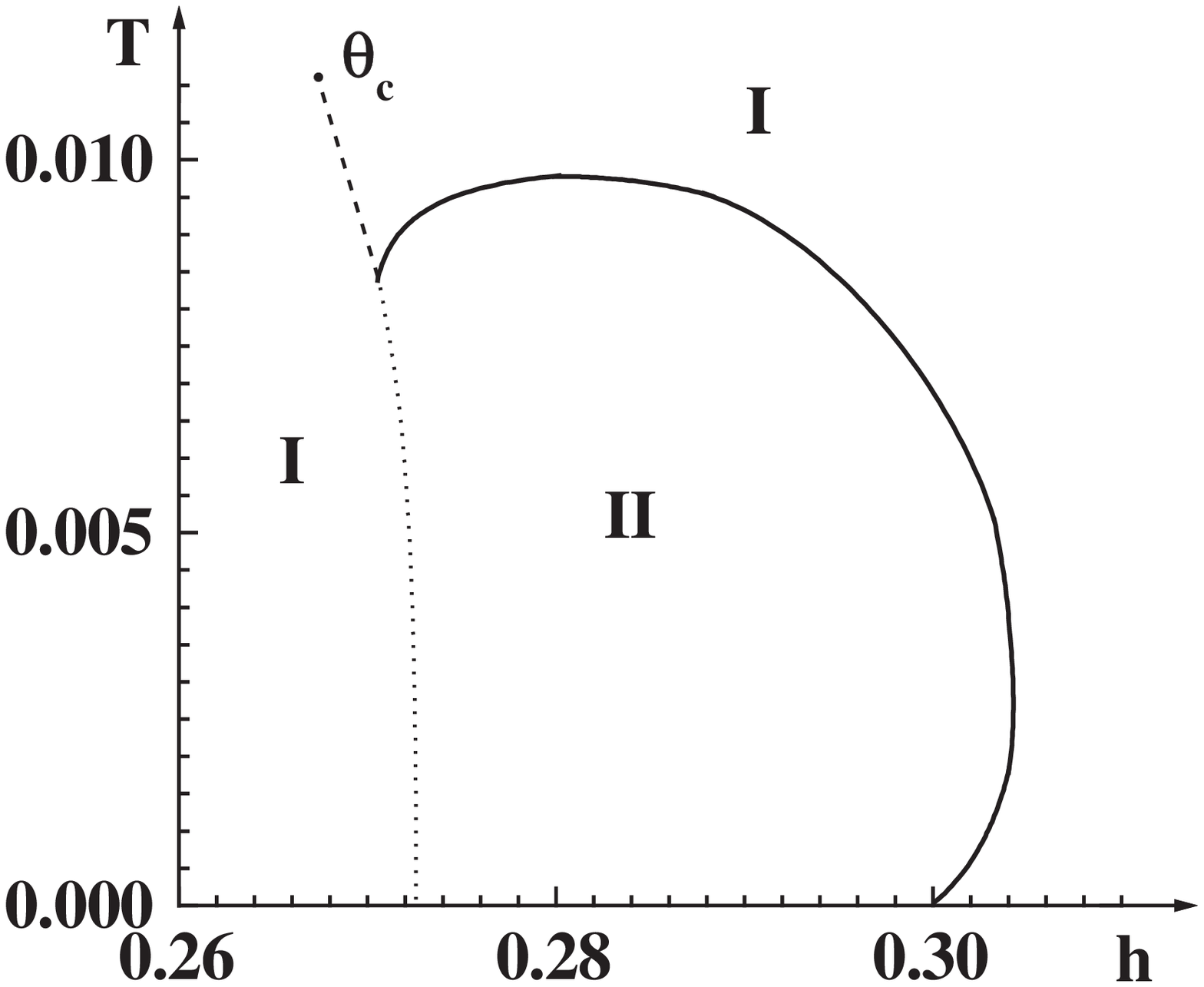}
\end{center}
 {\small  Figure~4. Phase diagram $T-h$ ($\mu =-0.36$, $W=0.2$, $g=1$).
 I -- uniform phase, II - chess-board phase.
 Dashed lines -- first order phase transitions between the
different uniform phases (bistability).
 Dotted line -- first order phase transitions between the
uniform and chess-board phases.
 Solid lines -- second order phase transitions.}
 $$\vspace{-.4cm}$$
 With the temperature increase the first order phase transition
between the uniform and chess-board phases transforms into the
first order phase transition between uniform phases and, finally,
disappear in the critical point {$\bf \theta_c$}.

 The diagram $T-h$ shows the possibility of the first order phase
transitions between uniform phases and either first or second
order ones between the uniform and chess-board phases at the
change of temperature.

 In figures~2 and 4 the case when the chemical potential is placed in the
lower band is presented.
 If the chemical potential is placed in the upper band our results
transform according to the symmetry of the Hamiltonian: $ \mu\to
-\mu$, $h\to 2g-h$, $n\to 2-n$, $S^z\to -S^z$.

\subsection{$n={\rm const}$ regime}

 In the regime of a fixed value of electron concentration the equilibrium
is defined by the minimum of free energy $F=\Omega +\mu N$.
 This condition form a set of equations (\ref{Szeq2}) and (\ref{neq})
for the pseudospin mean values and chemical potential.
 The obtained dependences of $F$ and $\mu$ on the electron concentration
are presented in figure~5.
\begin{center}
\raisebox{-2cm}[2cm][2cm]{\epsfysize 3.85cm\epsfbox{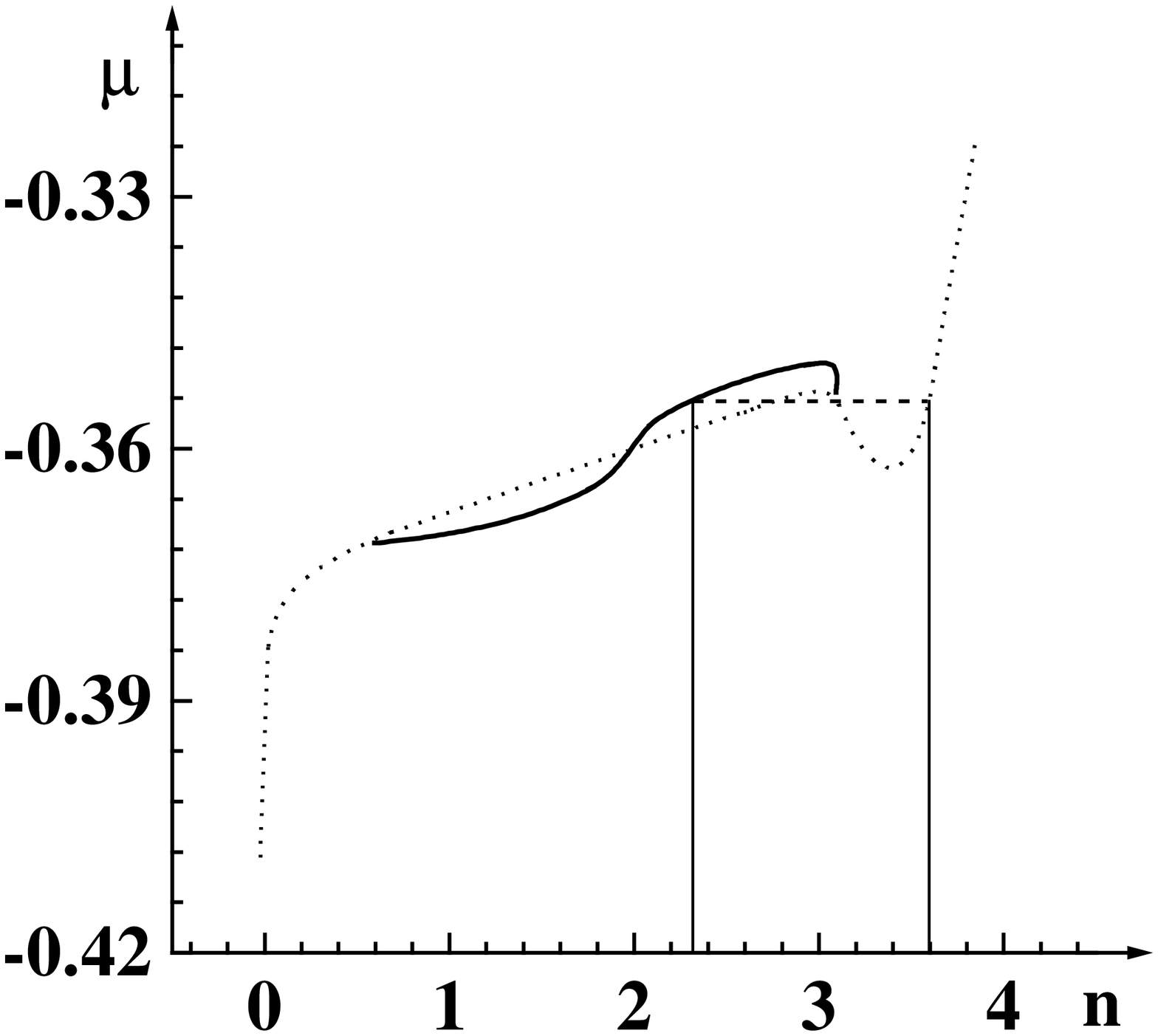}$\!$
                          \epsfysize 3.85cm\epsfbox{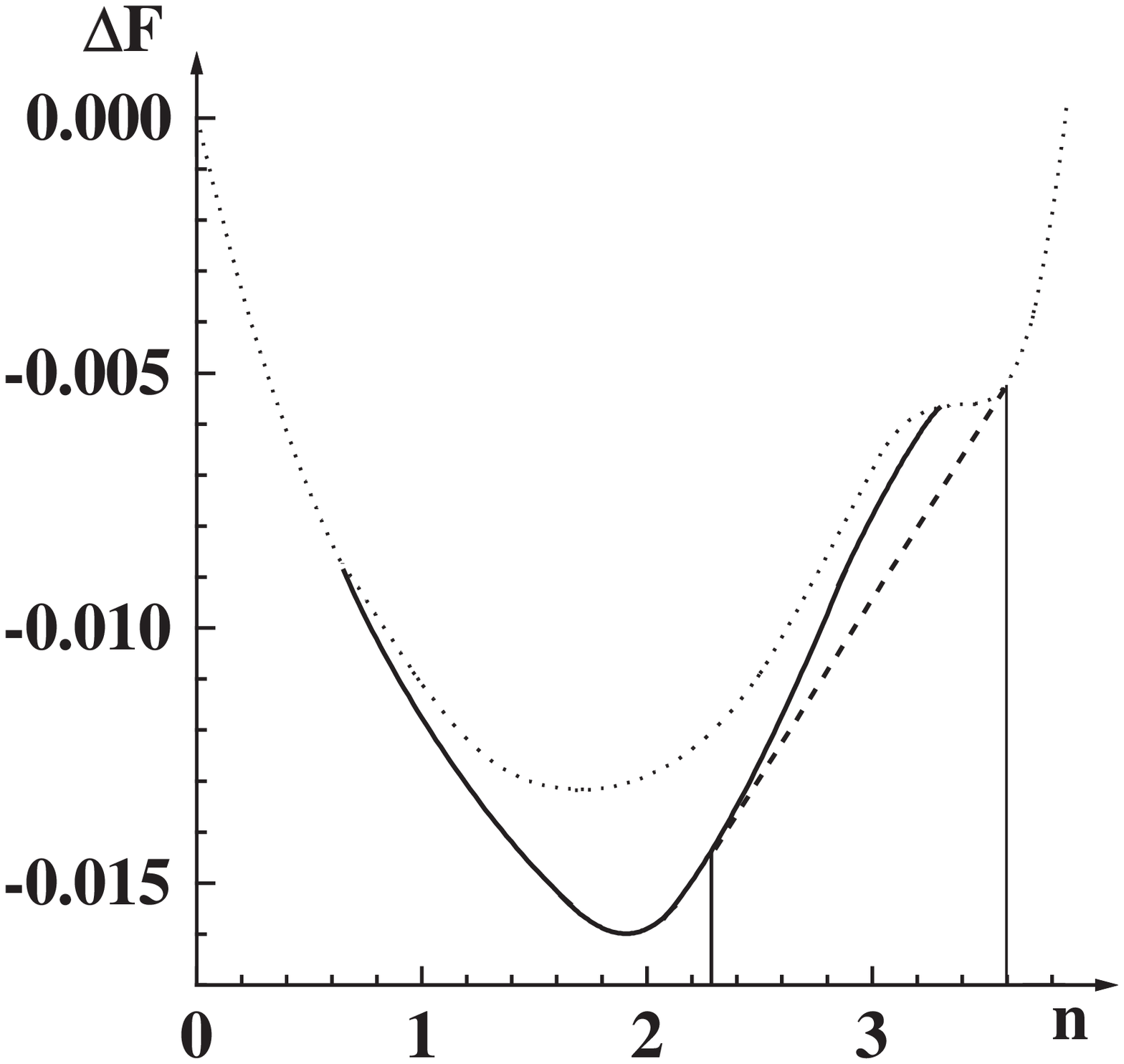}}
\end{center}
 {\small  Figure~5. Dependence of the chemical potential $\mu $ on the electron
concentration $n$ and deviation of the free energy from linear
dependence ($T=0.005$, $h=0.28$, $W=0.2$, $g=1$).
 Dashed line -- phase separation area.
 Dotted line -- uniform phase.
 Solid line -- chess-board phase.}
 $$\vspace{-.4cm}$$
 One can see the regions with ${\rm d}\mu /{\rm d}n \leqslant 0$
where states with a homogeneous distribution of particles are
unstable, which corresponds to the phase separation into the
regions with different phases (the uniform and chess-board ones in
this case) and with different electron concentrations and
pseudospin mean values (that is, different occupancies of particle
positions in the anharmonic potential wells).
 In the phase separated region the free energy, as
a function of $n$, deflects up and electron concentrations in the
separated phases are determined by the tangent line touch points.


 On the base of the obtained results the phase diagram $n-h$ was
constructed (figure~6).
\begin{center}
 \epsfxsize 0.4\textwidth\epsfbox{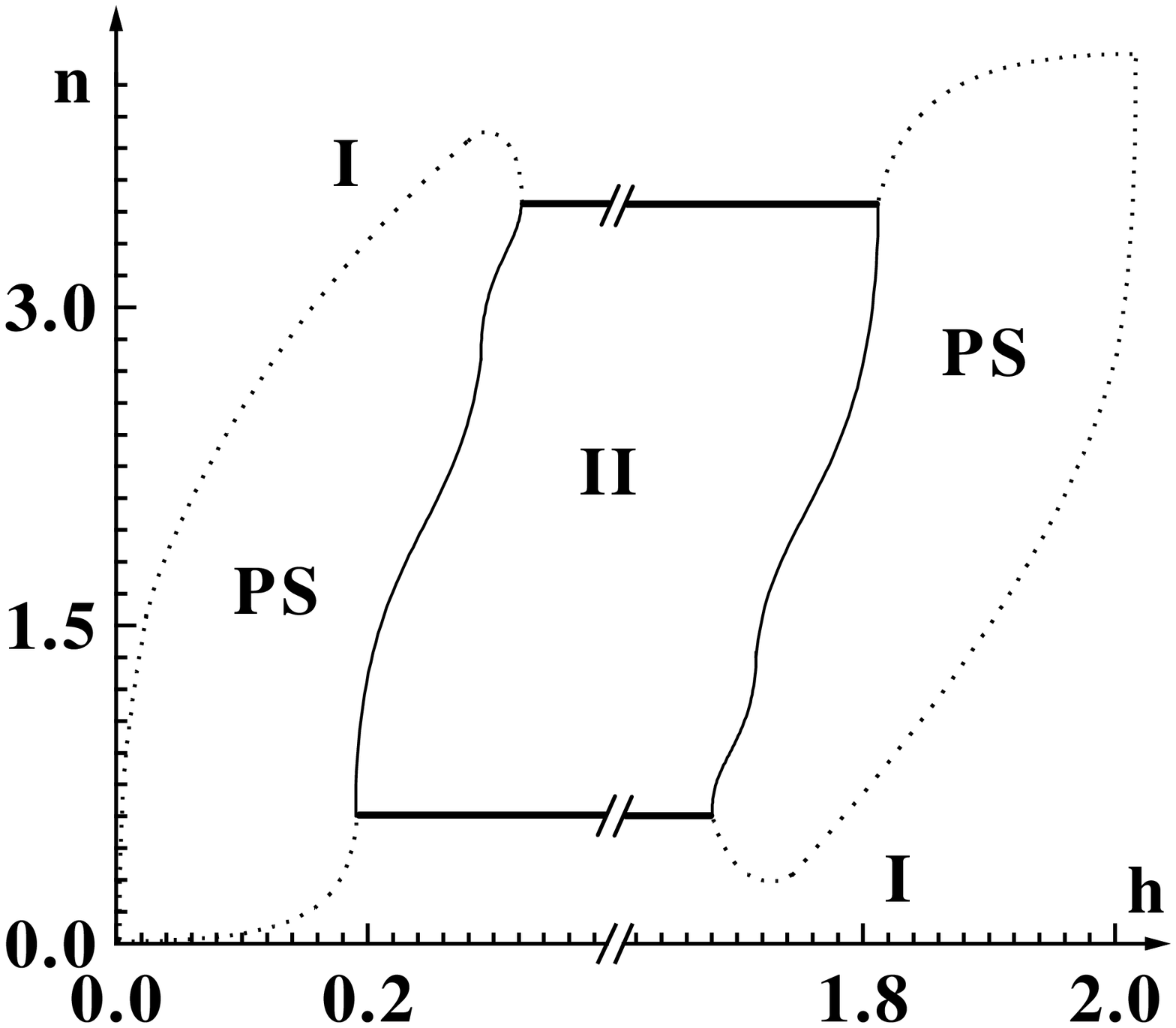}
\end{center}
 {\small  Figure~6. Phase diagram $n-h$ ($T=0.005$, $W=0.2$, $g=1$).
 I --  uniform phase,
 II -- chess-board phase,
 PS -- phase separation area.}
 $$\vspace{-.8cm}$$

 The  phase separation into the regions with uniform and
chess-board phases takes place in the case when chemical potential
is placed within the subbands $\varepsilon_2$, $\varepsilon_3$,
that agree with the results obtained in the $\mu={\rm const}$ case
when within this area we had the first order phase transition
between the corresponding phases.

 The phase separated and chess-board phase regions narrows with the
temperature increase, but thick solid lines in figure~6 approach
one to another faster and, for some temperatures, we have only the
phase separation into the regions with uniform phases.

\section{Conclusions}
 On the basis of the presented above self-consistent scheme for
calculation of the correlation and thermodynamical functions, the
energy spectrum, thermodynamics of phase transitions, possibility
of phase separations as well as appearance of the chess-board
phase have been investigated.
 The corresponding phase diagrams were build.
 Such comprehensive
  analysis of the thermodynamics of the considered
simplified PEM became possible due to the generalization of the
traditional GRPA approach (in which, on the basis of calculation
of the $\langle S^zS^z\rangle$ correlation function, only the
assumption about the chess-board phase appearance may be done).

 The obtained phase diagrams remind the situation know for the
Falicov-Kimball (FK) model (this model is close to PEM but differ
in the thermodynamic equilibrium conditions) with a rich phase
diagram.
 However, contrary to this model, an existence of the phase
transitions between uniform phases is possible in our case.
 This results from the another regime of thermodynamic averaging
(fixation of $h$ field which is an analogous to the chemical
potential for ions in the FK model).

 The study of the thermodynamics of the PEM within the framework
of the presented above approach provides reason enough to conclude
that on the basis of this model one can describe the phase
transitions and instabilities in the HTSC of the
YBa$_{2}$Cu$_{3}$O$_{7-\delta}$ type.
 Among them we should list the bistability effects (at the change
 of temperature), separation into phases with different electron
concentrations, pseudospin orientations (that corresponds to
different localization of particles in the anharmonic potential
wells).
 In general, it is in agreement with the picture observed experimentally
(some publications in this field were quoted above).
$$\vspace{-.2cm} $$
 1. Conradson S., Raistrick I.D. // Science, 1989, {\bf 243}, P.~1340.\\
 2. Mustre de Leon J., Conradson S.D. et al. // Phys. Rev. B., 1992, {\bf 45}, P.~2447.\\
 3. Mihajlovic D., Foster C.M. // Solid State Commun, 1990, {\bf 74}, P.~753.\\
 4. Ruani G., Taliami C. et al. // Physica C, 1994, {\bf 226}, P.~101.\\
 5. Iliev M.N., Hadjiev V.G., Ivanov V.G. // Journ. Raman Spectr., 1996, {\bf 27}, P.~333.\\
 6. Poulakis N., Palles D. et al. // Phys. Rev.~B, 1996, {\bf 53}, P.~R534.\\
 7. Testardi L.R., Moulton W,G. et al. // Phys. Rev. B, 1988, {\bf 37}, P.~2324.\\
 8. M\"uller V., Hucho C., Maurer D. // Ferroelectrics, 1992, {\bf 130}, P.~45--76.\\
 9. Mustre de Leon J., Batistic I. et al. // Phys. Rev. Lett., 1992, {\bf 68}, P.~3236.\\
 10. Ranninger J., Thibblin U. // Phys. Rev. B, 1992, {\bf 45}, P.~7730.\\
 11. Bishop A.R., Martin R.L., Muller K.A., Tesanovic Z. // Z.
 Phys. B -- Cond. Matter, 1989, {\bf 76}, P.~17.\\
 12. Gervais F. // Ferroelectrics, 1992, {\bf 130}, P.~117.\\
 13. Saiko A.P., Gusakov V.E. // JETP, 1995, {\bf 108}, P.~757.\\
 14. Cava R.J., Hewat A.W. // Physica C, 1990, {\bf 165}, P.~419.\\
 15. Bussman-Holder A., Simon A., Buttner H. // Phys. Rev. B, 1989,
 {\bf 39}, P.~207.\\
 16. Ranninger J. // Z. Phys. B, 1991, {\bf 84}, P.~167.\\
 17. Freericks J.K., Jarrell M., Mahan G.D. // Phys. Rev. Lett.,
 1996, {\bf 77}, P.~4588.\\
 18. M\"uller K.A. // Z. Phys. B -- Cond. Matter, 1990, {\bf 80}, P.~193.\\
 19. Hirsch J.E., Tang S. // Phys. Rev. B, 1989, {\bf 40},
 P.~2179.\\
 20. Frick M., von der Linden W., Morgenstern I., Raedt H.
 // Z. Phys. B -- Cond. Matter, 1990, {\bf 81}, P.~327.\\
 21. Stasyuk I.V., Shvaika A.M. Schachinger E. // Physica C, 1993,
 {\bf 213}, P.~57.\\
 22. Izyumov Yu.A., Letfulov B.M. // J. Phys.: Cond. Matter, 1990,
 {\bf 2}, P.~8905.\\
 23. Stasyuk I.V., Shvaika A.M. // Cond. Matt. Phys., 1994, No.~3,
 P.~134.\\
 24. Stasyuk I.V., Shvaika A.M., Danyliv O.D. // Molecular Phys.
 Reports, 1995, No.~9, P.~61.\\
 25. Stasyuk I.V., Havrylyuk Yu. // Cond. Matt. Phys., 1999,
 {\bf 2}, P.~487.\\
 26. Stasyuk I.V., Dublenych Yu. // Preprint of the Institute for
 Condensed Matter Physics, ICMP--99--07U, Lviv, 1999, 26~P. (in
 Ukrainian).\\
 27. Tabunshchyk K.V. // in: Ising Lectures--99, Preprint of the
 Institute for Condensed Matter Physics, ICMP--99--16, Lviv, 1999,
 P.~41--48\\
 28. Danyliv O.D., Stasyuk I.V. // Cond. Matt. Phys., 1996, No.~7,
 P.~163.\\
 29. Stasyuk I.V., Velychko O.V. // Ukrainian Journal of Physics, 1999,
 {\bf 44}, P.~772.\\
 30. Stasyuk I.V., Shvaika A.M., Tabunshchyk K.V. // Cond. Matt.
 Phys., 1999, {\bf 2}, P.~109.\\
 31. Stasyuk I.V., Shvaika A.M. // Journ. Phys. Studies, 1999, {\bf 3},
 P.~177.\\
 32. Stasyuk I.V., Shvaika A.M., Tabunshchyk K.V. // Acta Physica
 Polonica A, 2000, {\bf 97}, 1(2) (accepted for publication).\\
$$ $$
 Tabunshchyk K.V. e-mail: tkir@icmp.lviv.ua

\end{multicols}
\end{document}